# Radiation Pressure on Submerged Mirrors: Implications for the Momentum of Light in Dielectric Media

**Masud Mansuripur**

*College of Optical Sciences, The University of Arizona, Tucson, Arizona 85721*
*masud@optics.arizona.edu*

**Abstract**: Radiation pressure measurements on mirrors submerged in dielectric liquids have consistently shown an effective Minkowski momentum for the photons within the liquid. Using an exact theoretical calculation based on Maxwell's equations and the Lorentz law of force, we demonstrate that this result is a consequence of the fact that conventional mirrors impart, upon reflection, a 180° phase shift to the incident beam of light. If the mirror is designed to impart a different phase, then the effective momentum will turn out to be anywhere between the two extremes of the Minkowski and Abraham momenta. Since all values in the range between these two extremes are equally likely to be found in experiments, we argue that the photon momentum inside a dielectric host has the arithmetic mean value of the Abraham and Minkowski momenta.

## 1. Introduction

The 1953 experiments of Jones and Richards on submerged mirrors [1], followed by the refined 1977 experiments of Jones and Leslie [2], have shown beyond a doubt the dependence of the radiation pressure on the refractive index $n_o$ of the submerging liquid. Such findings, in turn, have been used to support the argument that the photons inside the liquid have the Minkowski momentum $n_o h f_o/c$, where $h$ is Planck's constant, $f_o$ is the light's frequency, and $c$ is the speed of light in vacuum [3, 4]. We show that the above result is a consequence of the fact that, for most mirrors, the Fresnel reflection coefficient at normal incidence, $\rho = |\rho| \exp(i\phi)$, has a phase $\phi \approx 180°$. If, however, $\phi$ is allowed to have other values, the radiation pressure on the submerged mirror will be reduced and, in particular, when $\phi$ approaches zero, the effective photon momentum will be found to reach the Abraham value of $hf_o/(n_o c)$. Our analysis thus suggests that, depending on the chosen value of $\phi$ for the mirror, the measured radiation pressure inside a dielectric medium would favor a photon momentum anywhere in the range between the Abraham and Minkowski values.

In section 2 we use the example of an idealized mirror, one with a negative value of the dielectric constant $\varepsilon$, to show that, when $n_o \neq 1$, the exactly calculated radiation pressure will have a strong dependence on $\phi$. Within the submerging liquid, an ideal flat mirror (i.e., one with 100% reflectance), sets up a perfect standing wave between the incident and reflected plane-waves. When $\phi = 180°$, the mirror's surface will be at the null point of the standing $E$-field; this is essentially the situation when conventional mirrors are used in the experiment, and the results of our calculations for the case of $\phi = 180°$ confirm the well-documented experimental findings [1, 2]. However, when $\phi$ deviates from 180°, the calculated radiation pressure drops; all else being the same, the ratio of the pressures on two mirrors, one with $\phi = 180°$ and the other with $\phi = 0°$, is found to be equal to $n_o^2$.

In section 3 we introduce a novel physical argument to demonstrate that the results of section 2 are *not* limited to certain (idealized) types of mirror, but are a general property of standing waves in a dielectric host. It will be shown that the radiation pressure, if observed locally within a standing wave, would be a function of location within the interference fringe. A submerged pressure sensor would thus detect either the Abraham or the Minkowski momentum depending on whether the sensor is located at the peak or the valley of an interference fringe. Since all locations within a fringe are equally accessible, the average photon momentum associated with a standing wave in a dielectric will thus have the arithmetic mean value of the Abraham and Minkowski momenta.

The argument of section 3 is employed in a different guise in section 4 in conjunction with a single plane-wave traveling in a dielectric host. The inescapable conclusion, once again, is that the plane-wave's momentum density is halfway between the Abraham and Minkowski values.

## 2. Radiation pressure on an ideal submerged mirror

The diagram in Fig. 1 depicts the interaction of light with a perfectly reflecting mirror whose index of refraction is the purely imaginary number $i n_1$ (i.e., the mirror's dielectric constant, $\varepsilon = -n_1^2$, is a negative real number). The incidence medium is a transparent dielectric of refractive index $n_o$. The normally incident plane-wave has frequency $f_o$, free-space wavelength $\lambda_o = c/f_o$, wave-number $k_z = n_o k_o = 2\pi n_o/\lambda_o$, and electromagnetic field amplitudes $(E_x, H_y) = (E_o, n_o E_o/Z_o)$, where $Z_o = (\mu_o/\varepsilon_o)^{1/2}$ is the impedance of the free space [5, 6].

The Fresnel reflection coefficient $\rho = (n_o - i n_1)/(n_o + i n_1)$ of the submerged mirror has unit magnitude for all values of $n_1$, but its phase angle, $\phi = -2 \arctan(n_1/n_o)$, can be anywhere in the range from 0° to 180° depending on the value of $n_1$. Beneath the surface of the mirror, the transmitted beam is an inhomogeneous plane-wave with an imaginary propagation vector $\boldsymbol{k} = i(2\pi n_1/\lambda_o)\hat{\boldsymbol{z}}$, which causes the beam amplitude to drop exponentially along the $z$-axis [6]. The transmitted $E$- and $H$-fields have a relative phase of 90°, yielding a time-averaged Poynting vector $<\boldsymbol{S}> = \tfrac{1}{2} \operatorname{Re}(\boldsymbol{E} \times \boldsymbol{H}^*) = 0$ everywhere inside the mirror; this, of course, is consistent with the mirror surface's 100% reflectance (i.e., $|\rho|^2 = 1$).

The radiation force per unit area of the mirror surface may be computed using the Lorentz force density $\boldsymbol{F} = \rho_b \boldsymbol{E} + \boldsymbol{J}_b \times \boldsymbol{B}$ exerted on the bound charge density $\rho_b = -\nabla \cdot \boldsymbol{P} = -\varepsilon_o(\varepsilon - 1)\nabla \cdot \boldsymbol{E} = 0$, and also on the bound current density $\boldsymbol{J}_b = \partial \boldsymbol{P}/\partial t = -i\omega \varepsilon_o(\varepsilon - 1)\boldsymbol{E}$ [7]. The force density $\boldsymbol{F}$, when integrated over the penetration depth of the light beam, yields

$$\begin{aligned}
<F_z> &= \int_0^\infty \tfrac{1}{2} \operatorname{Re}[-i\omega \varepsilon_o \mu_o (\varepsilon - 1) E_x(z) H_y^*(z)]\, dz \\
&= \tfrac{1}{2} \operatorname{Re}[i\omega \varepsilon_o \mu_o (n_1^2 + 1)(1 + \rho)(1 - \rho^*) n_o E_o^2/Z_o] \int_0^\infty \exp(-2 n_1 k_o z)\, dz \\
&= [n_o^2 (1 + n_1^2)/(n_o^2 + n_1^2)] \varepsilon_o E_o^2.
\end{aligned} \qquad (1)$$

Here $<F_z>$ is the time-averaged force per unit area of the mirror surface. For typical metallic mirrors, $n_1 \gg n_o$ and the above formula reduces to $<F_z> \approx \varepsilon_o n_o^2 E_o^2$. This is the expected result of assigning a Minkowski momentum density, $\boldsymbol{p}_M = \frac{1}{2}\,\text{Re}[\boldsymbol{D} \times \boldsymbol{B}^*]$, to the incident and reflected beams [7-10]. However, in the limit of small $n_1$, Eq. (1) yields $<F_z> \approx \varepsilon_o E_o^2$, which is consistent with the presence of the Abraham momentum density, $\boldsymbol{p}_A = \frac{1}{2}\,\text{Re}[\boldsymbol{E} \times \boldsymbol{H}^*]/c^2$, in the dielectric medium. Although ordinary metals at visible wavelengths have a large value of $n_1$, at higher frequencies (i.e., just below the plasma resonance frequency $\omega_p$ [5]) the metal's dielectric constant $\varepsilon$ assumes small negative values, leading to small (imaginary) values for the refractive index.

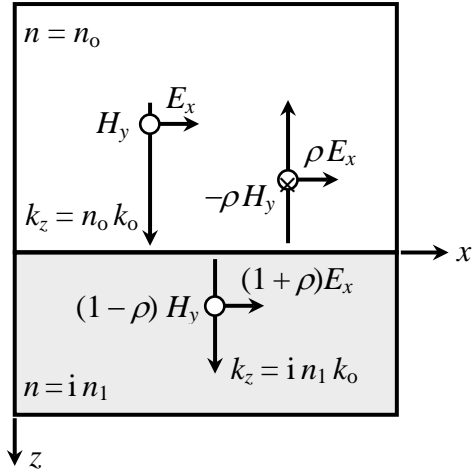

Fig. 1. Reflection of light from a mirror having a purely imaginary refractive index of i $n_1$. The incidence medium is a transparent dielectric of refractive index $n_o$. The normally incident plane wave has frequency $f_o$, free-space wavelength $\lambda_o = c/f_o$, wave-number $k_z = n_o k_o = 2\pi n_o/\lambda_o$, and field amplitudes $(E_x, H_y) = (E_o, n_o E_o/Z_o)$. The Fresnel reflection coefficient of the submerged mirror is $\rho = \exp(i\phi) = \exp[-2i\arctan(n_1/n_o)]$. Beneath the mirror's surface, the transmitted beam is an inhomogeneous plane-wave whose imaginary propagation vector $\boldsymbol{k} = i\,(2\pi n_1/\lambda_o)\,\hat{\boldsymbol{z}}$ causes the beam amplitude to drop exponentially along the $z$-axis.

We believe the above result is independent of the nature of the mirror, depending solely on the phase angle $\phi$ of the Fresnel coefficient. In fact, Eq. (1) may be written exclusively in terms of $\phi$ (with no explicit reference to the mirror's refractive index i$n_1$) as follows:

$$<F_z> = [1 + (n_o^2 - 1)\sin^2(\phi/2)]\varepsilon_o E_o^2. \tag{2}$$

Since multilayer dielectric mirrors having large reflectance and arbitrary values of $\phi$ can be readily designed, we expect the entire range of radiation pressures between $\varepsilon_o E_o^2$ and $\varepsilon_o n_o^2 E_o^2$ predicted by Eq. (2) to be amenable to experimental verification.

## 3. Electromagnetic momentum in standing waves

We present a general argument concerning the nature of the electromagnetic momentum in standing waves formed within transparent dielectric media. With reference to Fig. 2, two identical plane-waves propagating in opposite directions (along $\pm z$) create a standing-wave inside a dielectric medium of refractive index $n_o$. For each wave the $E$- and $H$-field amplitudes are $(E_x, H_y) = (E_o, n_o E_o/Z_o)$. The standing waves may thus be expressed as follows:

$$E_x(z, t) = 2E_o \sin(n_o k_o z)\sin(\omega t), \tag{3a}$$

$$H_y(z, t) = 2(n_o E_o/Z_o)\cos(n_o k_o z)\cos(\omega t). \tag{3b}$$

In Fig. 2 the sinusoidal curve having a period of $\lambda_o/2n_o$ depicts the intensity of the standing $E$-field, with the origin of the coordinate system chosen to coincide with one of its nulls. The standing $H$-field profile (not shown) is similar to that of the $E$-field, but shifted along the $z$-axis by $\lambda_o/4n_o$. The energy densities of the $E$- and $H$-fields, while stationary in space, oscillate in quadrature in time, so that the total optical energy swings back and forth between its electric and magnetic components.

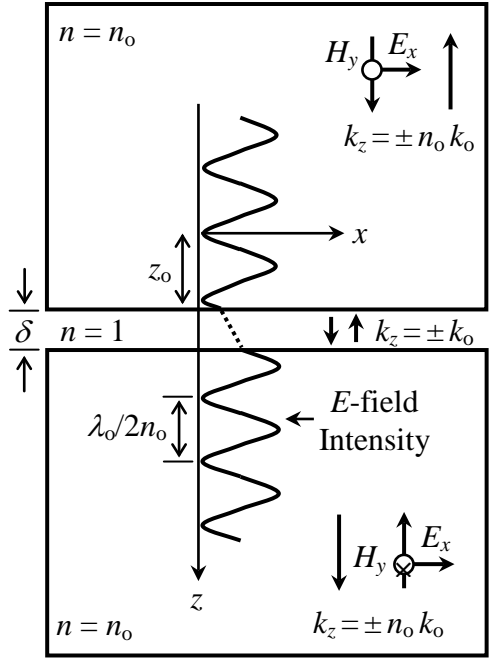

Fig. 2. Two identical plane-waves propagate in opposite directions (along $\pm z$) to create a standing wave inside a dielectric medium of refractive index $n_o$. The field amplitudes for each plane-wave are $(E_x, H_y) = (E_o, n_o E_o/Z_o)$. The sinusoidal curve having a period of $\lambda_o/2n_o$ depicts the intensity of the standing $E$-field, with the origin of the coordinate system chosen to coincide with one of its nulls; the standing $H$-field profile (not shown) is similar but shifted along the $z$-axis by $\lambda_o/4n_o$. Inside the narrow gap (width $\delta \ll \lambda_o$) located at $z = z_o$, the field amplitudes for each of the two counter-propagating plane-waves are $(E_x, H_y) = (E_g, E_g/Z_o)$.

An extremely narrow gap of width $\delta \ll \lambda_o$ is opened at $z = z_o$ to pierce into the medium in an attempt to discern the nature of the local $E$- and $H$-fields at that particular location within the standing wave. While the gap is too narrow to affect in any significant way the standing wave profiles, the $E$- and $H$-fields inside the gap differ substantially from those within the dielectric host. We denote by $E_g$ the $E$-field amplitude for each of the two counter-propagating plane-waves inside the gap; the corresponding $H$-field amplitude is then $H_y = E_g/Z_o$. The standing fields inside the gap are thus given by

$$E_x(z, t) = 2E_g \sin(k_o z + \psi) \sin(\omega t), \qquad (4a)$$

$$H_y(z, t) = 2(E_g/Z_o)\cos(k_o z + \psi) \cos(\omega t), \qquad (4b)$$

where $\psi$ is an as-yet-undetermined phase angle. The gap is sufficiently narrow that its upper and lower boundaries may be assumed to be effectively at the same location along the $z$-axis, namely, at $z = z_o$. Comparing Eqs. (3) and (4) shows that the continuity of $(E_x, H_y)$ at the gap boundaries requires the following identities:

$$E_o \sin(n_o k_o z_o) = E_g \sin(k_o z_o + \psi), \qquad (5a)$$

$$n_o E_o \cos(n_o k_o z_o) = E_g \cos(k_o z_o + \psi). \tag{5b}$$

These equations, when solved for the phase angle $\psi$ and the amplitude ratio $E_g/E_o$, yield

$$\tan(k_o z_o + \psi) = (1/n_o) \tan(n_o k_o z_o), \tag{6a}$$

$$E_g^2/E_o^2 = 1 + (n_o^2 - 1)\cos^2(n_o k_o z_o). \tag{6b}$$

Thus, according to Eq. (6b), if $z_o$ is varied from 0 to $\lambda_o/4n_o$, the value of $E_g$ would range from $n_o E_o$ to $E_o$. The gap field, of course, is the superposition of two identical counter-propagating beams in free space, each with its own (well-defined) momentum density $\pm\tfrac{1}{2}(\varepsilon_o E_g^2/c)\hat{z}$. At the null and peak positions of the $E$-field intensity within the dielectric, the plane-waves of the gap carry, respectively, the Minkowski and Abraham momenta of the dielectric medium. When the $E_g^2$ of Eq. (6b) is averaged over all values of $z_o$, each of the gap's plane-waves is seen to have an average momentum density equal to the arithmetic mean of the Abraham and Minkowski values associated with each plane-wave of the dielectric host. This is essentially the same conclusion as reached in our earlier papers [7, 11] via a different line of argument.

Returning now to the submerged mirror discussed in section 2, it must be clear that the phase $\phi$ of the reflection coefficient $\rho$ plays a role similar to that of $z_o$ in Eq. (6b). Thus when $\phi$ varies from 180° to 0°, it is as though the position of the mirror within the standing wave (produced by interference between incident and reflected beams) has shifted from $z_o = 0$ toward $z_o = \lambda_o/4n_o$. This is tantamount to substituting $\sin(\phi/2)$ for $\cos(n_o k_o z_o)$ in Eq. (6b), which would then reproduce the result in Eq. (2). The radiation pressure on the submerged mirror, derived in section 2 by a direct application of the Lorentz law of force, is now seen to be identical with the pressure exerted by the counter-propagating plane-waves within the (fictitious) gap. Needless to say, the gap must be at the same location relative to the standing wave of Fig. 2 as is the mirror surface with respect to the standing wave created by the superposition of the incident and reflected beams.

## 4. Momentum of a plane-wave

The introduction of a narrow gap inside a dielectric medium is a useful theoretical device that may be employed in an alternative way to provide further insight. Figure 3 shows a linearly polarized plane-wave propagating along the $z$-axis in a dielectric host of refractive index $n_o$; the field amplitudes inside the medium are $(E_x, H_y) = (E_o, H_o) = (E_o, n_o E_o/Z_o)$. Let us now imagine a narrow gap of width $\delta \ll \lambda_o$ in the host medium, and examine the nature of the fields inside this gap. The plane of the gap is $yz$ in (a) and $xz$ in (b). The electromagnetic field inside the gap is the superposition of two evanescent plane-waves, each of which must satisfy the constraints $\mathbf{k}\cdot\mathbf{k} = k_o^2$, $\mathbf{k}\cdot\mathbf{E} = 0$, and $(\mathbf{k}/k_o)\times\mathbf{E} = Z_o\mathbf{H}$ imposed by Maxwell's equations. The combined $E$- and $H$-fields of these evanescent waves must also satisfy the boundary conditions on both walls of the gap. In the case depicted in Fig. 3(a), continuity is required of the perpendicular $D$-field, $D_x = \varepsilon_o n_o^2 E_o$, and the tangential $H$-field, $H_y = H_o$, whereas in the case of Fig. 3(b) it is $E_x$ and $B_y = \mu_o H_y$ that must be continuous. In Fig. 3(a) the two (co-propagating) gap fields have the following $\mathbf{k}$-vector and field amplitudes:

$$\mathbf{k}_\pm/k_o = \pm i\sqrt{n_o^2 - 1}\,\hat{x} + n_o \hat{z} \tag{7a}$$

$$\mathbf{E}_\pm = \tfrac{1}{2}(n_o^2 \hat{x} \mp i n_o \sqrt{n_o^2 - 1}\,\hat{z}) E_o \tag{7b}$$

$$\mathbf{H}_\pm = \tfrac{1}{2}(n_o E_o/Z_o)\hat{y} = \tfrac{1}{2} H_o \hat{y}. \tag{7c}$$

It is readily verified that, in the limit when $\delta \to 0$, superposition of the above fields satisfies the continuity requirements at both boundaries. The momentum density $p$ in the gap is derived from the Poynting vector component $S_z$ along the propagation direction, namely,

$$p = <S_z>/c^2 = \tfrac{1}{2} \operatorname{Re}(2E_x \times 2H_y^*)/c^2 = \tfrac{1}{2} n_o^2 E_o H_o/c^2. \tag{8}$$

This is the Minkowski momentum of the plane-wave in the dielectric. A similar analysis for the case depicted in Fig. 3(b), where the plane of the gap is parallel to $xz$, yields

$$\boldsymbol{k}_\pm/k_o = \pm i\sqrt{n_o^2 - 1}\,\hat{\boldsymbol{y}} + n_o \hat{\boldsymbol{z}} \tag{9a}$$

$$\boldsymbol{E}_\pm = \tfrac{1}{2} E_o \hat{\boldsymbol{x}} \tag{9b}$$

$$\boldsymbol{H}_\pm = \tfrac{1}{2}(n_o \hat{\boldsymbol{y}} \mp i\sqrt{n_o^2 - 1}\,\hat{\boldsymbol{z}})\, E_o/Z_o = \tfrac{1}{2}(\hat{\boldsymbol{y}} \mp i\sqrt{1 - n_o^{-2}}\,\hat{\boldsymbol{z}})\, H_o. \tag{9c}$$

Here the momentum density in the gap is clearly that of Abraham, as it is given by

$$p = <S_z>/c^2 = \tfrac{1}{2} \operatorname{Re}(2E_x \times 2H_y^*)/c^2 = \tfrac{1}{2} E_o H_o/c^2. \tag{10}$$

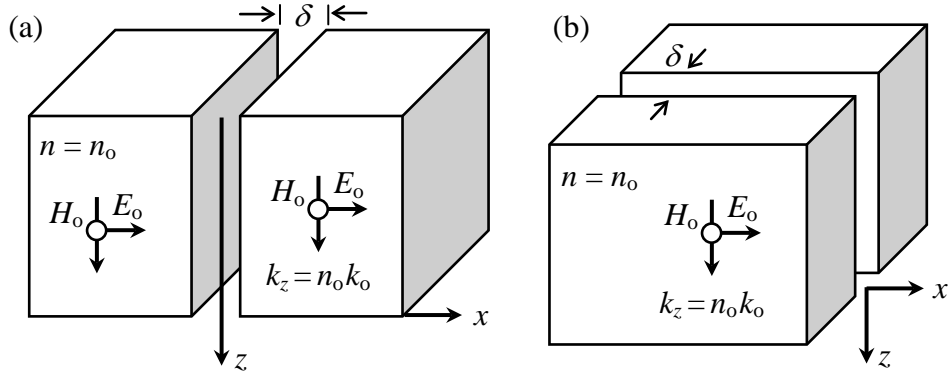

Fig. 3. A linearly polarized plane-wave propagates along the $z$-axis in a dielectric host of refractive index $n_o$; inside the medium, the field amplitudes are $(E_x, H_y) = (E_o, H_o)$. A narrow gap of width $\delta \ll \lambda_o$ is assumed to exist in this medium; the plane of the gap is $yz$ in (a) and $xz$ in (b). The electromagnetic field inside the gap is the superposition of two evanescent plane-waves whose combined fields satisfy the boundary conditions on both walls of the gap.

In the above example, the results depended on the choice of the gap orientation, as the symmetry of space was broken by the linear polarization of the beam. The field momentum, therefore, must be averaged over all possible gap orientations, leading to the mean value of the Abraham and Minkowski momenta. If, however, the beam happens to be circularly polarized, the results become independent of the gap orientation and yield the same mean value for the momentum density in all cases. For states of polarization other than circular, perhaps a better choice of the gap would be one in the form of a thin cylindrical shell of arbitrary radius, thickness $\delta \ll \lambda_o$, and cylinder axis aligned with the $z$-axis (i.e., the direction of propagation of the beam). All possible momentum densities will then occur at different locations around the circumference of this cylindrical shell, and the overall momentum density in the shell will coincide with the aforementioned average of the Minkowski and Abraham momenta.

### Acknowledgements

The author is grateful to Ewan Wright and Pavel Polynkin for helpful discussions. This work has been supported by the Air Force Office of Scientific Research (AFOSR) under contract number FA9550-04-1-0213.